\documentclass{elsart5p}

\usepackage{natbib}

\usepackage{amssymb}

\begin{document}

\begin{frontmatter}



\title{Giant Ly$\alpha$ nebulae in the high redshift ($z\ge$2) Universe}


\author{Montse Villar-Mart\'\i n$^1$}

\address{$^{1}$Instituto de Astrof\'\i sica de Andaluc\'\i a (CSIC), Aptdo. 3004, 18080 Granada, Spain
}

\begin{abstract}
High redshift radio galaxies ($z\ge$2) are believed to be progenitors of the
 giant ellipticals of today. They 
are often associated with giant Ly$\alpha$
nebulae (sometimes $>$100 kpc), which
             have been for more than two decades valuable sources 
of information
about the evolutionary status of the host galaxy and its chemical enrichment
and         star formation histories.

I present in this paper a summary of the most relevant results about
the giant nebulae obtained in the last $\sim$10 years 
and the implications on our understanding of the early
phases of evolution of massive elliptical galaxies.
An interesting earlier review  can be found in McCarthy (1993).

\end{abstract}





\end{frontmatter}


\label{}



\section{Introduction}

Our current belief is that the hosts of powerful radio sources in the distant
Universe are destined to become the giant ellipticals of today: the most
massive galactic systems we know (McLure et al. 1999).
High redshift radio galaxies (HzRG, $z>$2)  are, therefore, unique tools to investigate the early phases of massive
elliptical galaxies  in the high redshift Universe.

HzRG are often surrounded by giant Ly$\alpha$ nebulae which sometimes extend for more than 100 kpc (e.g.      Reuland et al. 2003, Villar-Mart\'\i n et al. 2003, McCarthy et al. 1990) and sometimes beyond the radio structures 
(e.g.   Maxfield et al. 2002, Pentericci et al. 2000, Eales et al. 1993).
Their morphologies are clumpy, irregular (with features such as filaments, plumes, ionization cones,
 e.g. Reuland et al. 2003) and often aligned with the radio axis 
(McCarthy, Spinrad \& van Breugel 1995). They are characterized by extreme kinematics, with measured FWHM  $\ge$1000 km s$^{-1}$ 
(e.g. Villar-Mart\'\i n et al. 2003, McCarthy, Baum \& Spinrad 1996), compared with values of a few hundred km s$^{-1}$ in low-redshift radio galaxies (e.g. Baum, Heckman \& van Breugel 1990, Tadhunter, Fosbury \& Quinn 1989). 

They have typical values of
ionized gas mass $\sim$10$^{9-10}$ M$_{\odot}$, Ly$\alpha$ luminosities
$\sim$several$\times$10$^{43-44}$ erg s$^{-1}$ and densities $n_e\sim$ few to several
hundred cm$^{-3}$ (e.g. McCarthy 1993, Villar-Mart\'\i n et al. 2003). 

They emit a rich  emission line spectrum (Fig.1) which reveals high levels
of metal enrichment  and excitation mechanisms 
mostly related to the nuclear activity (rather than stars), at least in the
direction along the
radio structures  (e.g. Vernet et al. 2001).

Such nebulae have been for more than two decades valuable sources 
of information
about the evolutionary status of the host galaxy, its chemical enrichment
history and the early phases of evolution of massive ellipticals.

\begin{figure}
\includegraphics{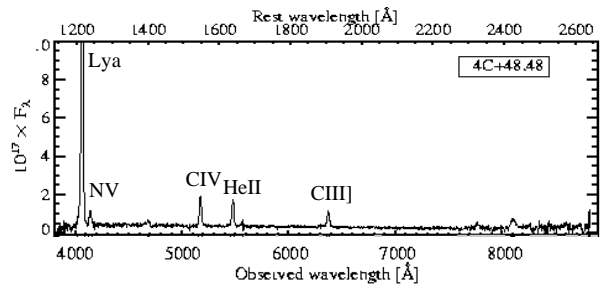}
\vspace{2.in}
\caption{Typical optical (rest frame UV) spectrum of a HzRG.
The main UV lines are indicated: Ly$\alpha$, NV$\lambda\lambda$1238,1242, CIV$\lambda\lambda$1548,1551, 
HeII$\lambda$1640 and CIII]$\lambda\lambda$1907,1909}
\end{figure}

In this review I present a summary of the most relevant
results  obtained in the last $\sim$10 years  about the giant nebulae
associated with HzRG. An interesting earlier review  can be found in McCarthy (1993).

\section{Jet gas interactions}

The  Ly$\alpha$ nebula that surround HzRG are in general aligned with the radio source axis 
(McCarthy, Spinrad \& van Breugel 1995). Their morphological and  kinematic properties 
(see \S1) are known to be strongly distorted by  interactions between the radio structures and the ambient
gas  (e.g.  Bicknell et al. 2000, van Ojik et al. 1997).

Detailed studies of low redshift powerful radio galaxies ($z<$0.5)  
have been very  useful to understand the 
 impact of such interactions  on the gaseous environment
(Tadhunter 2002, Villar-Mart\'\i n. et al. 1999, Clark et al. 1998, van Breugel
et al. 1985). At low redshift we have the advantage of a much
more detailed spatial information
and the possibility to use optical rest frame emission lines (the [OII]$\lambda$3727 doublet,
[OIII]$\lambda\lambda$4959,5007, H$\alpha$, etc)
 which are less affected by
reddening and, unlike some
strong UV rest frame  lines such as Ly$\alpha$ and CIV$\lambda\lambda$1548,1551, are not sensitive to resonance scattering
effects.

These studies have shown how jet-gas interactions perturb the kinematics, distort the morphology
of the nebula and change (decrease) the ionization level of the gas.
  We have identified  these effects in a sample of 10 radio galaxies at $z\sim$2-3. A good example is MRC~0943-242 at $z=$2.9 (see Fig. 2;  
Humphrey et al. 2006a, Villar-Mart\'\i n et al. 2003). This radio galaxy is associated with a giant nebula which extends for $\sim$70 kpc.
 The gas within the radio structures 
has higher surface brightness,  very perturbed kinematics 
(FWHM$>$1000 km s$^{-1}$) and lower ionization level. 
  The low surface brightness emission extending beyond the radio structures shows
quieter   kinematics (FWHM$\sim$400-600 km s$^{-1}$) and higher ionization level. These marked differences are a consequence of jet-gas interactions.

Although  evidence for jet-gas interactions has been found in many HzRG, 
the impact relative to other phenomena (such as pure  illumination by
the active nucleus (AGN) varies strongly from object to object. We have recently shown  (Humphrey et al. 2006a)
that the variation between objects on the ionization, kinematic and morphological properties of the 
giant nebulae associated with HzRG can be explained by the different impact of such interactions.
In general, small radio sources present more evidence for jet-gas interactions than large
radio sources (see also Best et al. 2000). Small radio sources are more likely to interact with the 
richer gaseous environment  which is likely to exist near the central AGN. 

\begin{figure}
\includegraphics{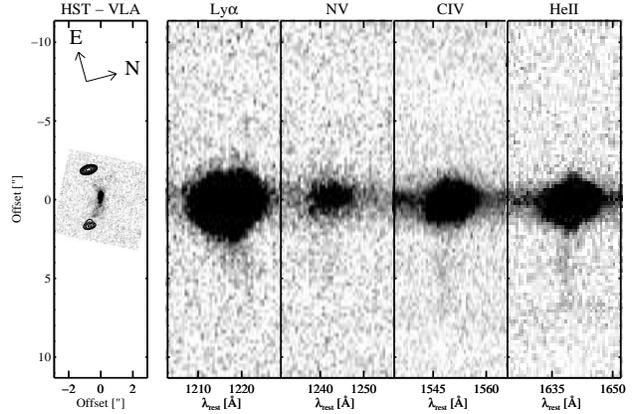}
\vspace{2.5in}
\caption{ MRC~0943-242 ($z=$2.9). 
  VLA radio
 map overlaid with the WFPC2 HST image (left), spatially aligned with the 2dim
Keck long slit spectra of the main UV rest frame emission lines.  The impact of jet-gas interactions on the observed
properties of the giant nebula ($\sim$70 kpc) is very clear in this radio galaxy. Notice the much broader and brighter emission lines within the radio
structures compared with the faint emission detected beyond.}
\end{figure}

Humphrey et al. (2006a) have found  evidence that the perturbed gas is part 
of a jet-induced {\it outflow}  whose effects can extend for tens of kpc from the nuclear region. This can have profound implications, since 
 such  mechanism 
may be more efficient than starburst
driven superwinds for massive galaxies to lose gas, eject metals  and  suppress star formation (Nesvabda et al. 2006; see also Rawlings 2003).
It could be an effective mechanism for regulating galaxy growth and polluting the
intergalactic medium with metals. This issue deserves  and will surely be subject
of  further
investigation.

\section{Metal enrichment}

The emission line spectrum of ionized nebulae 
 contains valuable information about the chemical abundances
of the gas and, ultimately,  the  history of formation of the stars
responsible for producing the metals. Numerous studies of low $z$ objects
have proven the power of such technique, based on optical rest
frame emission lines such
as [OIII]$\lambda\lambda$4959,5007, H$\alpha$, [NII]$\lambda\lambda$6548,6583, [SII]$\lambda\lambda$6716,6731, etc.

For HzRG, this issue is complicated for two main reasons:
until recently, the chemical abundances of the gas were investigated
using  the UV rest-frame lines, redshifted into the optical window. The behaviour of some of these lines
 (e.g. Ly$\alpha$, NV$\lambda\lambda$1238,1242, CIV$\lambda\lambda$1548,1551) is very complex due to resonance scattering
effects and the much higher sensitivity to dust.

On the other hand, there is an important complication derived from
the uncertainty on what  the dominant ionizing mechanism of the
gas is: is it
jet-driven shocks or photoionization by the active nucleus? (see Tadhunter
2002 for a review). The degeneracy between shock and AGN photoionization model predictions is
the main reason why this issue has not been disentangled after many years
of discussion.

There are a few clear cut cases, such as the radio galaxy  MRC~0943-242 at $z=$2.9, discussed in \S2 
(Villar-Mart\'\i n et al. 2003). 
 The uncertainty
of what ionizes the gas dissapears in  this case:
 the quiescent gas (i.e. not perturbed by jet-gas interactions) which extends beyond the radio structures (Fig.~2) must be clearly
photoionized. The strength of the NV emission relative
to the other  lines such as CIV and HeII (Fig.~3) implies at least half solar abundances (Humphrey
2005, Humphrey et al. 2007, in prep.).

Such cases are exceptional, however, and in general the emission from the
shocked, perturbed gas and the quiescent gas are blended.
Several years ago, by modeling the rich UV emission line spectra 
of a sample of nine HzRG   ($z\sim$2-3),
 we found that the EELRs of  these objects are characterized by solar and supersolar metallicities. Our work was, however, seriously limited by the fact that  we ignored  shocks  and we assumed the same 
ionization level for all objects (Vernet et al. 2001). 

\begin{figure}
\includegraphics{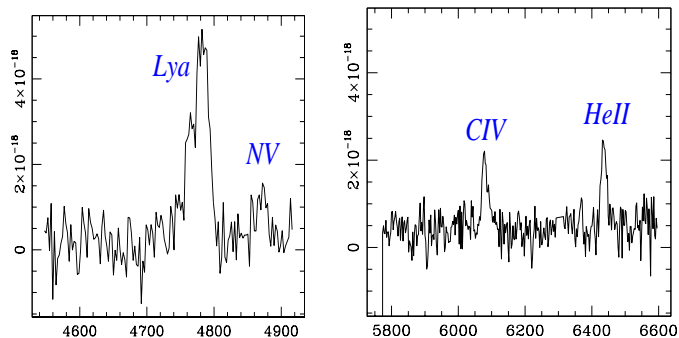}
\vspace{2.in}
\caption{Spectrum of the quiescent photoionized nebula associate with the radio galaxy MRC~0943-242 at
$z=$2.9. The relative strength of the NV emission implies at least half
solar abundances. Flux in units of erg s$^{-1}$ cm$^{-1}$ \AA$^{-1}$.}
\end{figure}

During the last few years we have improved  this work substantially (Humphrey
2005, Humphrey et al. 2007, in prep.) by: 1) using both shocks and
AGN photoionization models 2) investigating the impact of varying different
parameters such as density, $U$, power law spectral index, metallicity 3) enlarging
the data set with optical rest frame emission lines, 
including of course,  traditional lines such as [OIII], H$\beta$, H$\alpha$, [NII], [SII], etc.
The resulting rest frame far UV-optical spectra have a coverage of $\sim$35 emission lines  that is 
unprecedented for radio galaxies at any redshift.

One of the main conclusions of this work  is that,
as in low $z$ radio galaxies (Robinson et al. 1987),
the data are well explained by photoionization with $U$ varying between
objects (Fig.~4). In general, photoionization
is the main ionization mechanism, with shock collisional ionization having a minor role
for only a few emission lines.
Whether the
dominant source of photoionization is the active nucleus or the continuum
generated by the shocked gas cannot always be said, although the kinematic
and radio/optical morphological properties give additional valuable information
about the relative impact of shocks (Humphey et al. 2006a).

The critical issue here is that,  whatever the ionizing continuum source is
(AGN vs. hot shocked gas),
if we use those emission lines whose excitation is dominated by photoionization,
we can then
 put constraints on the metal abundances using standard
techniques based on the modeling of the emission line spectrum of
photoionized nebulae.

\begin{figure*}
\includegraphics{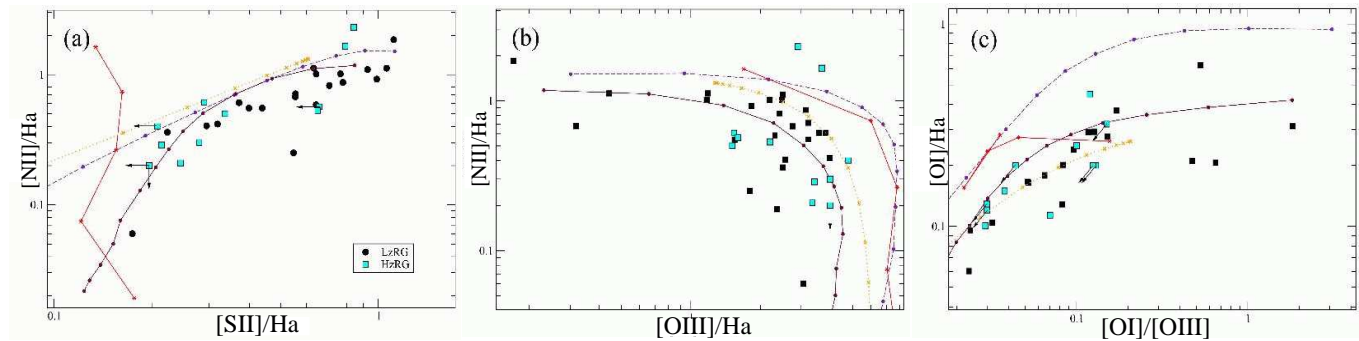}
\vspace{2.6in}
\caption{Comparison between our sample of HzRG (light blue symbols) and the low z radio galaxy sample (black symbols)
of Robinson et al. (1987). Both samples overlap in the diagrams. This suggests that the ionized gas in both samples is characterized
by similar abundances (within a factor of two solar) and physical properties. 
The colour lines are predictions of different models. We will highlight here the dark purple sequence, which
best reproduces the data. Notice that the data set seems
 to follow this one parameter sequence characterized by a standard power law 
ionizing continuum
with index $\alpha$=-1.5, solar abundances and such that  the ionization parameter $U$ varies along the sequence. Taken from Humphrey (2005).} 
\end{figure*}

With this work we confirm  the conclusion of Vernet et al. (2001) that
the chemical abundances of the EELR of powerful HzRG
are close to their solar values and the metallicity
does not vary by more than a factor of two in our sample. We reject, on
the other hand, the interpretation of a metallicity gradient between
objects.

At low $z$, this type of studies (e.g. Robinson et al. 1987) have shown that the
EELRs of powerful radio galaxies are characterized by  abundances of at least 1/10 solar, and
they are probably within a factor of 2 of solar. We have plotted (Fig.~4) our high $z$ radio galaxy sample
in three diagnostic diagrams together with the low $z$ sample of Robinson et al. (1987).
The diagrams show that there is no significant difference between the two samples, suggesting
that they are also similar in metal abundances and physical conditions of the ionized nebulae.

 The high chemical abundances in the giant nebulae associated
with HzRG suggest that
such objects have already undergone  intense star formation activity. It is consistent
with chemical evolution models for giant ellipticals
 and supports the idea that distant powerful radio galaxies are progenitors of the most
massive galaxies we know 
(Vernet et al. 2001).

\section{Star formation in HzRG and the giant Ly$\alpha$ nebulae}

Investigating the existence of young stars {\it in} HzRG (rather than in
companion objects, e.g. Pentericci et al. 2000, Venemans et al. 2005) using optical data has
proven to be a difficult task. 
 Because of  the powerful active
nucleus, the optical (UV rest frame) continuum is likely to be contaminated, sometimes dominated, 
by components related to the nuclear activity, i.e., nebular continuum and/or scattered light. Polarization and emission line measurements
are essential to determine whether   a young stellar population also contributes 
(e.g. Vernet et al. 2001,  Cimatti et al. 1998, Dey et al. 1997). Submm studies have been
more efficient on the search for young stars in HzRG (e.g. Stevens et al. 2003, Archibald et al. 2001). Provided that the dust  responsible for the submm emission
is heated by young stars (e.g. Tadhunter et al. 2005)  enormous star formation rates  of
thousands  M$_{\odot}$ yr$^{-1}$ are implied.

In spite of the difficulties inherent to optical studies,
the existence of young stars in HzRG has been inferred
indirectly from polarization studies (e.g. Vernet et al. 2001, Cimatti et al. 1998, Dey et al. 1997).
Although scattered light contaminates the UV rest frame continuum in numerous radio galaxies,
the level of polarization changes noticeably between objects. This is more naturally explained
if a young stellar population dilutes the polarization.

So, the presence of young stars in a fraction of HzRG is relatively well established.
What has not been clear so far  is whether they contribute to the excitation of
 the giant Ly$\alpha$ nebulae.
Using  their emission line spectra 
to investigate this issue
 (as in non active star forming galaxies) might seem impossible, since the
excitation of the gas  is predominantly a consequence of AGN related processes. However, recent results suggest that young stars do 
contribute to the excitation of the Ly$\alpha$ nebula in some HzRG.

We have recently found (Villar-Mart\'\i n et al. 2006a) 
that $\sim$54\% of HzRG at z$\ge$3 and $\sim$8\% of radio galaxies
at 2$\le z<$3 emit Ly$\alpha$ unusually strong compared with the general population of HzRG.  The Ly$\alpha$/HeII and Ly$\alpha$/CIV ratios and the Ly$\alpha$ luminosities
in these objects (that we have called Ly$\alpha$ excess objects or LAEs) are $\sim$2-3.5 times higher than in the vast majority of HzRG (see also
De Breuck et al. 2000).
Star formation  is the most successful explanation
to supply the extra continuum photons needed to explain the Ly$\alpha$ excess.
Star forming rates of $\sim$several hundred M$_{\odot}$ yr$^{-1}$ are implied
by our models (ignoring Ly$\alpha$ absorption and dust reddening). This interpretation is strongly supported 
 by the clear trend for objects
with lower optical continuum polarization level to show larger Ly$\alpha$/HeII ratios
(see Fig. 5). It is possible that LAEs have recently
undergone a merger event which triggered the formation of young stars and the AGN and radio activities.

 Our interpretation is further
supported by the tentative trend found
by other authors for $z\ge$3 radio galaxies to show lower UV-rest frame polarization levels (De Breuck et al. 2007, in prep.)
or the dramatic increase on the detection rate at submm wavelengths of $z>$2.5 radio galaxies (Archibald et al. 2001).

We argue that, although the fraction of LAEs may be incompletely determined, 
both at 2$\le z<3$ and at $z\ge3$,  the much larger fraction of LAEs found at $z\ge$3 is  a genuine redshift evolution
and not due to selection effects (Villar-Mart\'\i n et al. 2006a). Therefore, 
  our study suggests that the radio galaxy phenomenon was
 more often associated with a massive starburst at $z>$3 than at $z<$3.  In other words,  powerful radio galaxies (and, according to the unification model, 
powerful radio quasars),  appear in more actively star forming galaxies at $z>$3 than at $z<$3.

\begin{figure}
\includegraphics{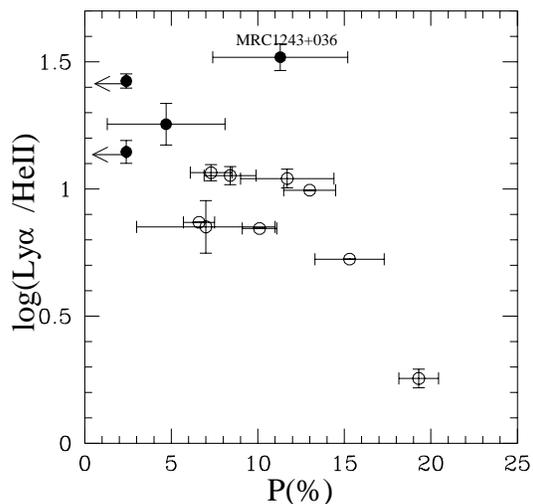}
\vspace{2.7in}
\caption{log(Ly$\alpha$/HeII) vs. percentage of polarization of the optical
continuum (UV rest frame) for all radio galaxies at $z\ge$2 for which
both have been measured. Black solid symbols correspond to LAEs, 
while hollow symbols are  objects with no Ly$\alpha$ excess.
Errorbars are shown when available. Notice the clear trend for objects with the highest Ly$\alpha$/HeII ratios to show lower polarization, as expected if the Ly$\alpha$ enhancement
is due to stellar photoionization. The trend is broken by 1243+036, which
shows too high $P(\%)$ for its large Ly$\alpha$/HeII ratio (but see Villar-Mart\'\i n et al. 2006a).}
\end{figure}

\section{The origin}

Different scenarios have been proposed to explain the origin of the giant nebulae
associated with HzRG. It has been suggested that they could be cooling flow nebulae or
gas ejected by jet-, AGN- or starburst-driven superwinds or giant rotating structures
(e.g. Villar-Mart\'\i n et al. 2003,2006b; van Ojik et al. 1996). 

The kinematic pattern of the nebulae could provide important information on this regard.
However, the nuclear activity complicates our work again. As explained in \S2, 
the jet-induced
shocks distort the gas kinematics, which does not reflect the intrinsic original pattern. 

Around the year 2000, most spectroscopic studies of HzRG had been based on long slit data
with the slit aligned along the radio structures, where the impact of jet-gas interactions
is likely to be strongest. Van Ojik et al.  published an interesting
paper in 1996 on the radio galaxy MRC~1243+036 ($z=$3.6). In addition to the typical
kinematically perturbed gas ($\ge$1000 km s$^{-1}$) often found in HzRG, the authors detected a low surface brightness
 giant nebula  
($\sim$140 kpc) which extends well beyond the radio structures and with very quiescent
kinematics (FWHM and velocity shift $\sim$few hundred km s$^{-1}$). It is clear that this gas is not affected by the interactions with
the radio structures.

In spite of the profound impact that the jet-induced shocks  have on the nebular
properties,
 we expect naturally, that there are large regions of the giant nebulae which do not notice the
passage of the radio structures. The results of van Ojik et al. (1996)
suggested to us that it should be possible to detect the non-perturbed gas.
If we could isolate the emission from this ionized 
 gas in other objects, we would be in an excellent position to study the ionization,
morphological and kinematic properties of the nebulae, previous to any jet-induced distortion.
Ultimately, this information could be critical to understand the origin of the nebulae and its possible link with the formation process of the galaxy.
  
So, we looked for the quiescent  nebulae in a sample of 10 HzRG ($\sim$2-3),
based on long slit spectroscopic data obtained with the LRISp spectrograph on Keck II (see
Vernet et al. 2001 for a description of the observations and the data).
We found that, in addition to the highly perturbed gas,
all objects are   embedded in giant (often $\ge$100
kpc), low surface brightness   nebulae  of metal rich, ionized gas
with {\it quiescent kinematics}, with FWHM and velocity shifts of $\sim$several hundred km s$^{-1}$  
(Villar-Mart\'\i n et al. 2002, 2003).

With the goal of investigating the kinematic, ionizaton and morphological properties of
the nebulae in two spatial dimensions, we started an observational program of 3D integral field spectroscopy with the fiber-fed 
integral field spectrographs Vimos on VLT
and PMAS/PPAK on the 3.5m telescope in Calar Alto.

I show in Fig.~6 some results on one of the 6 radio galaxies we have observed. It is the
 velocity and dispersion fields of the {\it quiescent} giant Ly$\alpha$ nebula 
($\sim$120 kpc)
associated with the radio galaxy MRC~2104-242 ($z=$2.49) based on VIMOS data
(Villar-Mart\'\i n et al. 2006b). Such 
an ordered and symmetric kinematic pattern is quite striking for a HzRG.
Our first thought was that we were looking at a giant rotating structure,
whose motions imply a  dynamical mass of $\ge$3$\times$10$^{11}$ M$_{\odot}$. 
However, 
we found that, due to the lack of knowledge on   the intrinsic gas distribution (the nebula is
 anisotropically illuminated by the central active nucleus), we could not discriminate between
rotation and radial motions. In the latter case, at least we could
reject outflows, based on  the  radio properties of the source.

Based on the thesis work of Andy Humphrey (2005), we have advanced 
substantially 
on  our understanding of the
intrinsic kinematic pattern of the giant quiescent nebulae (Humphrey et al. 2006b).
Making use of the LRISp long-slit spectra of the Keck sample discussed above, and the valuable information provided
by the radio maps (Carilli et al. 1997), we have identified several correlated asymmetries: on the side of the
brightest radio jet and hot spot (i)  the quiescent nebulae have the highest
redshift, (ii) Ly$\alpha$ is brighter relative to the other lines and continuum,
(iii) the radio spectrum is flattests and (iv) the radio structure has its highest
polarization. Interestingly, the correlation (i) also appears to be present 
in powerful radio galaxies with 0$<z\le$1.

Collectively, these asymmetries are most naturally explained as an effect
of orientation, with the quiescent nebulae in infall (Fig.~6).  This is the first
study to distinguish between the rotation, infall, outflow and chaotic motion
scenarios for the kinematically quiescent emission line nebulae around 
 powerful AGN.

The infalling gas of HzRG is highly enriched with metals (see \S3) and
these objects are likely to have undergone already an intense period of star
formation activity (e.g. Villar-Mart\'\i n et al. 2006b). Therefore, it is not certain
that the infall process is related to the early phases of formation of the
central galaxy. Since we find that
infall seems to be be taking place in  powerful radio galaxies at low $z$ as well,
 where the elliptical galaxy is already highly evolved, it is clear that infall
does not necessarily imply an early phase in the formation process.

\begin{figure}
\includegraphics{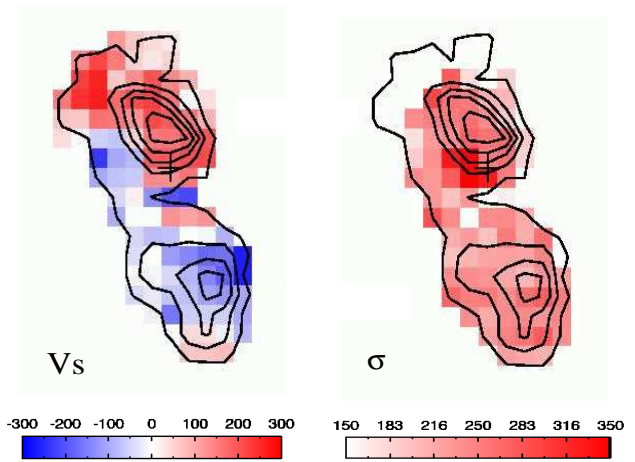}
\vspace{2.6in}
\caption{MRC~2104-242. Vimos  velocity (left) and dispersion (right) fields of the giant
Ly$\alpha$ nebula ($\sim$120 kpc) associated with the radio galaxy MRC~2104-242 ($z=$2.49). The position of the radio core is indicated with a cross.
The velocity field appears symmetric and ordered implying either rotation
or radial motions. }
\end{figure}

\begin{figure*}
\includegraphics{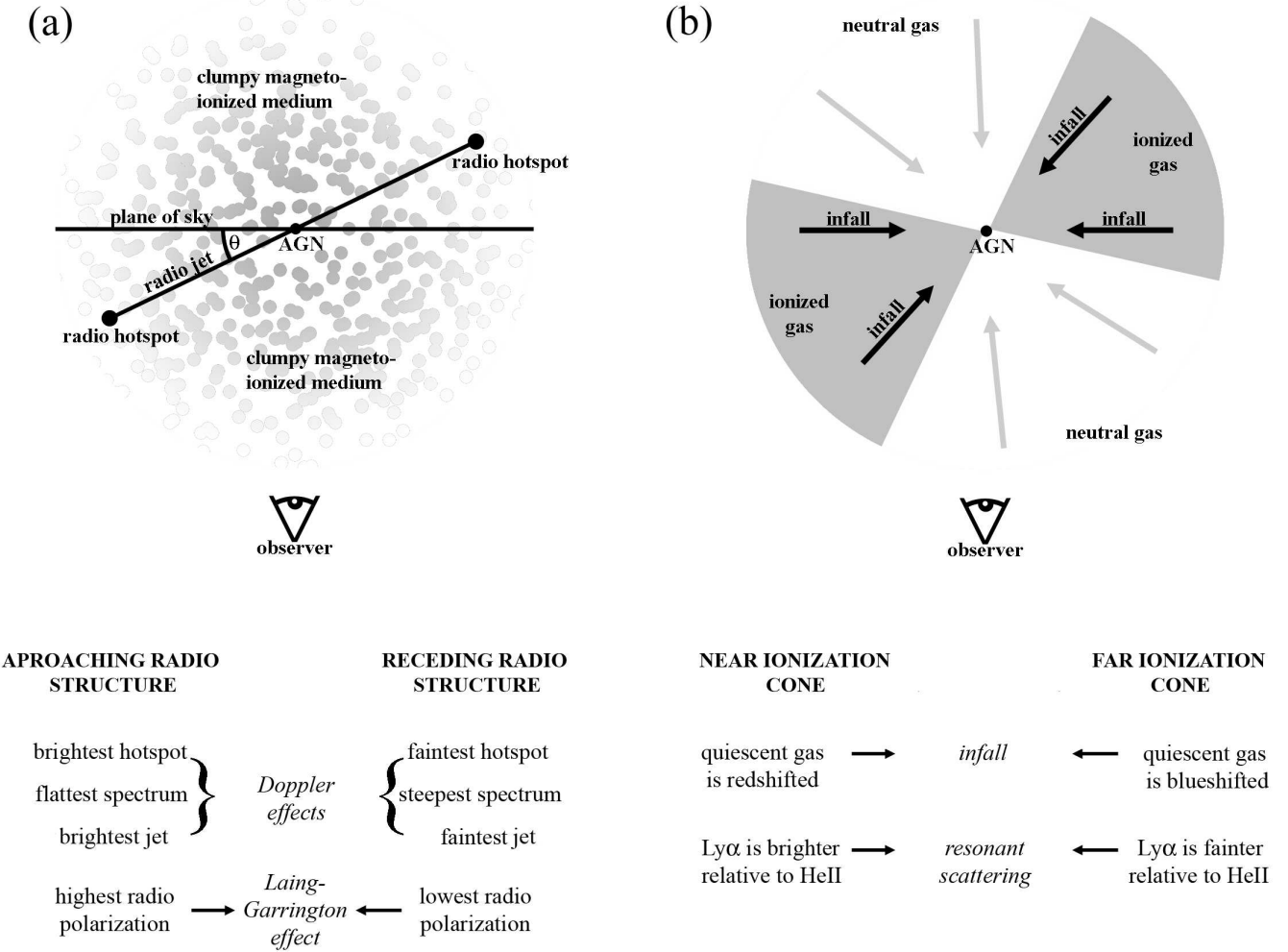}
\vspace{5in}
\caption{Cartoon to illustrate our infall scenario. (a) Radio source asymmetries. If the radio structure is viewed at an angle, then as a result of Doppler shifting and boosting the approaching radio hotspot is observed to be brighter,
and to have a flatter radio spectrum. In comparison to the receding radio structure, the approaching radio structure is viewed through a lower column of the ambient magneto-ionic depolarizing medium, and consequently is observed to have a higher radio polarization (i.e. is less depolarized); this is analogous to the Laing-Garrington effect. (b) UV-optical emission line asymmetries. The ionization bicone shares a common axis with the radio structure, and thus is also viewed at an angle. The near cone is associated with the approaching radio structure and the far cone is associated with the receding radio structure. The kinematically quiescent gas is infalling towards the nucleus, and thus the near cone is observed to be redshifted (in projection) relative to the far cone. The space beyond the ionization bicone contains neutral Hydrogen and, due to the orientation of the bicone/radio axis, the near cone is viewed through a lower column of neutral Hydrogen. Consequently, the Ly$\alpha$ emission from the near cone suffers less resonant scattering that that from the far cone. Taken from Humphrey et al. 2006b.}
\end{figure*}

\section{Summary and conclusions}

High $z$ radio galaxies ($z>$2) are associated with ionized nebulae
which often extend for more than 100 kpc. The kinematic, ionization
and morphological properties of the nebulae are strongly influenced
by the nuclear activity. 

The giant nebulae consist of two main gaseous components:   perturbed
and quiescent gas. The first component is usually
spatially located within the radio structures while the quiescent
gas often extends beyond. The perturbed gas has lower ionization level
and extreme kinematics (FWHM of the lines $>$1000 km $^{-1}$ vs. several hundred
 km $^{-1}$ for the quiescent gas). Such marked differences between
both gaseous components are naturally explained by the impact of jet-gas interactions. 

While the perturbed gas is part of a jet-induced outflow, the
 quiescent component, which has not been affected by the passage
of the radio structures, is infalling towards the center. 
The giant nebulae are characterized by high metal abundances (at least half solar).
Although AGN related processes dominate the ionization mechanism
of the gas, evidence for stellar photoionization has been found in some
radio galaxies (most of them at $z>$3). 

At $z<$3, powerful radio galaxies
 have already undergone an intense period of star formation activity
and chemical enrichment, which explains the high metal abundances.
Different studies suggest that   powerful radio galaxies at $z>$3
were more often associated with a massive starburst.

A plausible scenario to describe  HzRG is this: these massive
galaxies formed the bulk of
their stars at $z>$3 (e.g. Seymour et al. 2006, Villar-Mart\'\i n et al. 2006b), embedded in the giant infalling nebulae. 

Since then, the nebulae have undergone a rapid chemical evolution
process. At some point, the
nuclear activity starts and the central quasar acts like a flash bulb that
illuminates the gaseous environment rendering the nebulae visible at
tens of kpc from the nuclear region.

When the radio activity is triggered and the radio structures
advance trough the nebula, the shocked gas reverses its 
kinematic pattern and a powerful
outflow is generated which accelerates the gas at high speeds,
in some cases higher than the escape velocity of the system.
The rest of the nebula keeps falling towards the center. 

Since
infall is found in radio galaxies at all $z$, and the available
gas mass could be consumed in less that 10$^8$ yr, an interesting possibility
is that the radio activity is fed by the infalling gas so
that it is only detected when the infall is happening and feeding
efficiently the active nucleus.

A similar variety of Ly$\alpha$ morphologies, sizes, line luminosities and
 FWHM values characteristic of high $z$ radio galaxies
has been observed in high redshift {\it radio quiet} Ly$\alpha$ nebulae
(e.g. Ly$\alpha$ blobs, Steidel et al. 2000, van Breugel et al. 2006).
 Although in these objects the radio structures are not present,
what has been learned for more than 20 years about giant nebulae
associated with high $z$ radio galaxies, can be very valuable to
understand the nature and origin of the recently discovered radio
quiet Ly$\alpha$ nebulae.

\section*{Acknowledgments}
Thanks to Tom Oosterloo, and very specially Raffaella Morganti, 
for organizing this interesting, productive and enjoyable meeting.
Thanks to Andy Humphrey for useful comments on the manuscript and
 to my
collaborators, L. Binette, M. Cohen, C. De Breuck, B. Fosbury, A. Humphrey,
S. S\'anchez, S. di Serego Alighieri and J. Vernet  for their contribution to this project.

\end{document}